\documentclass[twocolumn]{webofc}
\usepackage[varg]{txfonts}   
\begin{document}
\title{Final results for the neutron
       $\beta$-asymmetry parameter $A_0$ from the \\ UCNA experiment}
%
%

\author{
\firstname{B.} \lastname{Plaster}\inst{1}\fnsep
\thanks{\email{brad.plaster@uky.edu}} \and
\firstname{E.} \lastname{Adamek}\inst{2} \and
\firstname{B.} \lastname{Allgeier}\inst{1} \and
\firstname{J.} \lastname{Anaya}\inst{3} \and
\firstname{H.O.} \lastname{Back}\inst{4,5} \and
\firstname{Y.} \lastname{Bagdasarova}\inst{3,6} \and
\firstname{D.B.} \lastname{Berguno}\inst{7} \and
\firstname{M.} \lastname{Blatnik}\inst{8} \and
\firstname{J.G.} \lastname{Boissevain}\inst{3} \and
\firstname{T.J.} \lastname{Bowles}\inst{3} \and
\firstname{L.J.} \lastname{Broussard}\inst{5,9} \and
\firstname{M.\ A.-P.} \lastname{Brown}\inst{1} \and
\firstname{R.} \lastname{Carr}\inst{8} \and
\firstname{D.J.} \lastname{Clark}\inst{3} \and
\firstname{S.} \lastname{Clayton}\inst{3} \and
\firstname{C.} \lastname{Cude-Woods}\inst{4} \and
\firstname{S.} \lastname{Currie}\inst{3} \and
\firstname{E.B.} \lastname{Dees}\inst{4,5} \and
\firstname{X.} \lastname{Ding}\inst{7} \and
\firstname{S.} \lastname{Du}\inst{4} \and
\firstname{B.W.} \lastname{Filippone}\inst{8} \and
\firstname{A.} \lastname{Garc\'{i}a}\inst{6} \and
\firstname{P.} \lastname{Geltenbort}\inst{10} \and
\firstname{S.} \lastname{Hasan}\inst{1} \and
\firstname{A.} \lastname{Hawari}\inst{4} \and
\firstname{K.P.} \lastname{Hickerson}\inst{8} \and
\firstname{R.} \lastname{Hill}\inst{3} \and
\firstname{M.} \lastname{Hino}\inst{11} \and
\firstname{J.} \lastname{Hoagland}\inst{4} \and
\firstname{S.A.} \lastname{Hoedl}\inst{6,12} \and
\firstname{G.E.} \lastname{Hogan}\inst{3} \and
\firstname{B.} \lastname{Hona}\inst{1} \and
\firstname{R.} \lastname{Hong}\inst{6} \and
\firstname{A.T.} \lastname{Holley}\inst{4} \and
\firstname{T.M.} \lastname{Ito}\inst{3} \and
\firstname{T.} \lastname{Kawai}\inst{13} \and
\firstname{K.} \lastname{Kirch}\inst{3} \and
\firstname{S.} \lastname{Kitagaki}\inst{14} \and
\firstname{A.} \lastname{Knecht}\inst{6} \and
\firstname{S.K.} \lastname{Lamoreaux}\inst{3} \and
\firstname{C.-Y.} \lastname{Liu}\inst{2} \and
\firstname{J.} \lastname{Liu}\inst{8,15} \and
\firstname{M.} \lastname{Makela}\inst{3} \and
\firstname{R.R.} \lastname{Mammei}\inst{7} \and
\firstname{J.W.} \lastname{Martin}\inst{16} \and
\firstname{N.} \lastname{Meier}\inst{4} \and
\firstname{D.} \lastname{Melconian}\inst{6,17} \and
\firstname{M.P.} \lastname{Mendenhall}\inst{8} \and
\firstname{S.D.} \lastname{Moore}\inst{4} \and
\firstname{C.L.} \lastname{Morris}\inst{3} \and
\firstname{R.} \lastname{Mortensen}\inst{3} \and
\firstname{S.} \lastname{Nepal}\inst{1} \and
\firstname{N.} \lastname{Nouri}\inst{1} \and
\firstname{R.W.} \lastname{Pattie, Jr.}\inst{4,5} \and
\firstname{A.} \lastname{P\'{e}rez~Galv\'{a}n}\inst{8} \and
\firstname{D.G.} \lastname{Phillips, II}\inst{4} \and
\firstname{A.} \lastname{Pichlmaier}\inst{3} \and
\firstname{R.} \lastname{Picker}\inst{8} \and
\firstname{M.L.} \lastname{Pitt}\inst{7} \and
\firstname{J.C.} \lastname{Ramsey}\inst{3} \and
\firstname{R.} \lastname{Rios}\inst{3,18} \and
\firstname{R.} \lastname{Russell}\inst{8} \and
\firstname{K.} \lastname{Sabourov}\inst{4} \and
\firstname{A.L.} \lastname{Sallaska}\inst{6} \and
\firstname{D.J.} \lastname{Salvat}\inst{2} \and
\firstname{A.} \lastname{Saunders}\inst{3} \and
\firstname{R.} \lastname{Schmid}\inst{8} \and
\firstname{S.J.} \lastname{Seestrom}\inst{3} \and
\firstname{C.} \lastname{Servicky}\inst{4} \and
\firstname{E.I.} \lastname{Sharapov}\inst{19} \and
\firstname{S.K.L.} \lastname{Sjue}\inst{3,6} \and
\firstname{S.} \lastname{Slutsky}\inst{8} \and
\firstname{D.} \lastname{Smith}\inst{4} \and
\firstname{W.E.} \lastname{Sondheim}\inst{3} \and
\firstname{X.} \lastname{Sun}\inst{8} \and
\firstname{C.} \lastname{Swank}\inst{8} \and
\firstname{G.} \lastname{Swift}\inst{5} \and
\firstname{E.} \lastname{Tatar}\inst{18} \and
\firstname{W.} \lastname{Teasdale}\inst{3} \and
\firstname{C.} \lastname{Terai}\inst{4} \and
\firstname{B.} \lastname{Tipton}\inst{8} \and
\firstname{M.} \lastname{Utsuro}\inst{13} \and
\firstname{R.B.} \lastname{Vogelaar}\inst{7} \and
\firstname{B.} \lastname{VornDick}\inst{4} \and
\firstname{Z.} \lastname{Wang}\inst{3} \and
\firstname{B.} \lastname{Wehring}\inst{4} \and
\firstname{J.} \lastname{Wexler}\inst{4} \and
\firstname{T.} \lastname{Womack}\inst{3} \and
\firstname{C.} \lastname{Wrede}\inst{6} \and
\firstname{Y.P.} \lastname{Xu}\inst{4} \and
\firstname{H.} \lastname{Yan}\inst{1} \and
\firstname{A.R.} \lastname{Young}\inst{4,5} \and
\firstname{J.} \lastname{Yuan}\inst{8} \and
\firstname{B.A.} \lastname{Zeck}\inst{3,4}
}

\institute{
University of Kentucky, Lexington, Kentucky 40506, USA 
\and
Indiana University, Bloomington, Indiana 47408, USA 
\and
Los Alamos National Laboratory, Los Alamos, New Mexico 87545, USA 
\and
North Carolina State University, Raleigh, North Carolina 27695, USA 
\and
Triangle Universities Nuclear Laboratory, Durham, North Carolina 27708, USA 
\and
University of Washington, Seattle, Washington 98195, USA 
\and
Virginia Tech, Blacksburg, Virginia 24061, USA 
\and
California Institute of Technology, Pasadena, California 91125, USA 
\and
Duke University, Durham, North Carolina 27708, USA 
\and
Institut Laue-Langevin, 38042 Grenoble Cedex 9, France 
\and
Kyoto University, Kumatori, Osaka, 590-0401, Japan 
\and
Princeton University, Princeton, New Jersey, 08544, USA 
\and
Kyoto University, Kumatori, Osaka, 590-0401, Japan 
\and
Tohoku University, Sendai 980-8578, Japan 
\and
Shanghai Jiao Tong University, Shanghai, 200240, China 
\and
University of Winnipeg, Winnipeg, MB R3B 2E9, Canada 
\and
Texas A\&M University, College Station, Texas 77843, USA 
\and
Idaho State University, Pocatello, Idaho 83209, USA 
\and
Joint Institute for Nuclear Research, 141980 Dubna, Russia 
}

\abstract{The UCNA experiment was designed to measure the neutron
$\beta$-asymmetry parameter $A_0$ using polarized ultracold neutrons
(UCN).  UCN produced via downscattering in solid deuterium were
polarized via transport through a 7 T magnetic field, and then
directed to a 1 T solenoidal electron spectrometer, where the decay
electrons were detected in electron detector packages located on the
two ends of the spectrometer.  A value for $A_0$ was then extracted
from the asymmetry in the numbers of counts in the two detector
packages.  We summarize all of the results from the UCNA experiment,
obtained during run periods in 2007, 2008--2009, 2010, and 2011--2013,
which ultimately culminated in a 0.67\% precision result for $A_0$.}
\maketitle
\section{Introduction}
\label{intro}
Precision measurements of neutron $\beta$-decay observables, together
with precise Standard Model calculations, constitute a sensitive test
for new physics \cite{gonzalez-alonso-ppns}.  The UCNA experiment
\cite{pattie09, liu10, plaster12, mendenhall13,
brown18} determined the neutron $\beta$-asymmetry parameter $A$, the
angular correlation between the neutron's spin and the decay
electron's momentum, which appears in the angular distribution of the
emitted electrons as \cite{jackson57}
\begin{equation}
d\Gamma(E_e, \theta) \propto 1 + P A \beta \cos\theta.
\end{equation}
Here, $E_e$ denotes the electron's energy, $\beta = v/c$ where $v$ is
the electron's velocity, $P$ is the neutron polarization, and $\theta$
is the angle between the neutron's spin and the electron's momentum.
At lowest order, measurements of $A$ determine the ratio of the
weak axial-vector and vector coupling constants, $\lambda \equiv g_A/g_V$,
according to \cite{jackson57}
\begin{equation}
A_0 = -2 \frac{\lambda^2-|\lambda|}{1 + 3\lambda^2}.
\end{equation}

The UCNA experiment was carried out at the Ultracold Neutron Facility
at the Los Alamos Neutron Science Center \cite{saunders13, ito18}, and
was the first-ever measurement of any neutron $\beta$-decay angular
correlation parameter using Ultracold Neutrons (UCN).  UCNA has
provided for the determination of $A$ via a complementary technique to
cold neutron beam-based measurements of $A$, such as from the PERKEO
III experiment \cite{saul-ppns, markisch18}, via the use of different techniques
for the neutron polarization, different sensitivity to environmental
and neutron-generated backgrounds, and different methods for electron
detection, among others.

\section{Overview of the UCNA Experiment}
\label{overview}

\begin{figure}[h]
\centering
\includegraphics[scale=0.35]{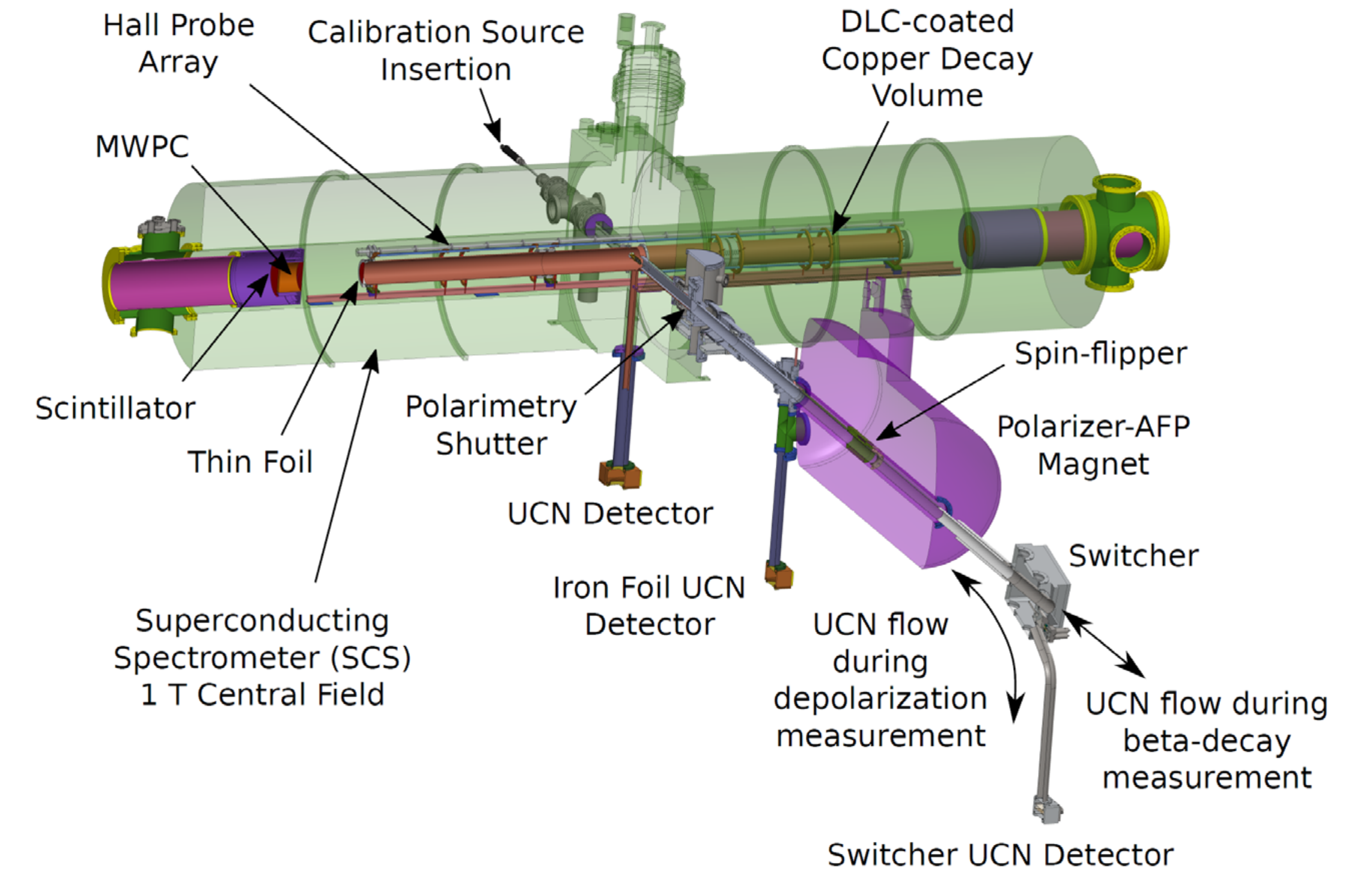}
\caption{Schematic diagram showing the primary components of the UCNA
experiment, including the 7 T polarizing magnet, the spin flipper,
the electron spectrometer, and the UCN detector at the switcher (used
for polarization measurements).}
\label{fig1}
\end{figure}

An overview of the basic operating principles of the UCNA experiment
\cite{plaster12} is as follows, of which a schematic diagram is shown
in Fig.  \ref{fig1}.  A pulsed 800 MeV proton beam, with a
time-averaged current of 10 $\mu$A, was incident on a tungsten
spallation target.  The emerging neutrons were moderated in cold
polyethylene, then downscattered to the ultracold regime in a crystal
of solid deuterium.  A so-called ``flapper valve'', located above the
solid deuterium crystal, opened after each proton beam pulse, allowing
the UCN to escape, and then closed soon afterwards, to minimize UCN
losses in the deuterium.

After emerging from the source, the UCN were transported along a
series of guides through a polarizing solenoidal magnet
\cite{holley12} where a 7 T peak field provided for spin state
selection (by rejecting the low-field seeking spin state).
Immediately downstream of the 7 T peak field, the polarizing magnet
was designed to have a low-field-gradient 1 T region, along which a
birdcage-style adiabatic fast passage (AFP) spin-flipper resonator
\cite{holley12} was located.  The spin-flipper provided the ability to
flip the spin of the neutrons presented to the electron spectrometer,
important for minimization of various systematic effects in the
measurement of the asymmetry.

The polarized UCN that emerged from the polarizer and the AFP
spin-flipper region were then transported to a 1 T solenoidal
spectrometer \cite{plaster08}, where a 3-m long cylindrical decay trap
was situated along the spectrometer's axis.  There, the UCN spins were
aligned parallel or anti-parallel to the magnetic field direction, and
the emitted decay electrons then spiraled along the field lines
towards one of two electron detector packages located on the two ends
of the spectrometer, providing for the measurement of the
asymmetry from the rates of detected electrons in the two detector
packages.

When the spectrometer magnet was commissioned in the mid-2000's, the
central 1 T field region was uniform to the level of $\pm 3 \times
10^{-4}$ over the length of the UCN decay trap \cite{plaster08}.
However, over time, due to damage to the magnet's shim coils (as a
result of numerous magnet quenches), the field uniformity was somewhat
degraded, resulting in a $\sim 30$~Gauss ``field dip'' near the center
of the decay trap region \cite{plaster12}.  One important feature of
the spectrometer's field profile is that the field was expanded, such
that the UCN decays occurred in the 1 T region, but the electron
detectors were located in a 0.6 T field region, which minimized
Coulomb backscattering and other effects related to the measurement of
the asymmetry.

A little more detail on the asymmetry measurement in the electron
spectrometer is as follows.  The two electron detector packages
consisted of multiwire proportional chambers (MWPCs) \cite{ito07},
backed by a plastic scintillator disk \cite{plaster08}.  The MWPCs,
with their orthogonally-oriented cathode planes, provided for a
measurement of the center position of the spiraling electron
trajectory in both transverse directions, which permitted
reconstruction of the transverse coordinates of where the electron
originated within the UCN decay volume, important for the definition
of a fiducial volume.  Light from the plastic scintillator was
transported along a series of light guides to four photomultiplier
tubes (PMTs).  The light from the scintillator provided for a
measurement of the decay electron's energy, and the timing from the
scintillators provided for a relative determination of the electron's
initial direction of incidence (in the event the electron
backscattered in such a way that it was detected in both scintillator
detectors).

It is important to point out that the decay electrons necessarily
traversed a number of thin foils between the decay trap and the
electron detector packages.  In particular, the ends of the decay trap
were sealed off with thin foils, the purpose of which was to increase
the UCN density in the decay trap, thus increasing the detected rate
of neutron decays.  Then, the MWPC fill gas (100 Torr of neopentane)
was sealed off from the spectrometer vacuum by thin entrance and exit
foils.

The thickness of these foils over the course of the running of the
experiment, from 2007--2013, is summarized in Table \ref{tab:foils}.
I will emphasize that the experiment evolved from operation in 2007
with decay trap foils consisting of 2.5 $\mu$m thick Mylar coated with
0.3 $\mu$m of Be and 25 $\mu$m thick Mylar MWPC foils, to its final
configuration in 2013, in which the decay trap foils were reduced to
0.15 $\mu$m thick 6F6F coils \cite{hoedl03} coated with 0.15 $\mu$m of
Be and 6 $\mu$m thick Mylar MWPC foils.

\begin{table}[h]
\caption{Summary of the decay trap and MWPC foil thicknesses over the
course of the running of the UCNA experiment.  Prior to the 2012--2013
data taking run, the decay trap foils consisted of Mylar coated with
Be (such that, e.g., ``2.5 + 0.3'' indicates 2.5 $\mu$m of Mylar
coated with 0.3 $\mu$m of Be).  For the 2012--2013 data taking run, the
decay trap foils consisted of 0.15 $\mu$m thick 6F6F foils
\cite{hoedl03} coated with 0.15 $\mu$m of Be.  For all years, the MWPC
foils were composed of Mylar.}  \centering
\begin{tabular}{lcc} \hline\hline
& Decay Trap& MWPC \\
Data Set& Foils [$\mu$m]& Foils [$\mu$m] \\ \hline
2007 \cite{pattie09}& 2.5 + 0.3& 25 \\
2008--2009 \cite{liu10, plaster12}& 0.7 + 0.2 | 13.0 + 0.2& 6 | 25 \\
2010 \cite{mendenhall13}& 0.7 + 0.2& 6 \\
2011--2012 \cite{brown18}& 0.50 + 0.15& 6 \\
2012--2013 \cite{brown18}& 0.15 + 0.15& 6 \\ \hline\hline
\end{tabular}
\label{tab:foils}
\end{table}

\section{Polarization}
\label{polarization}

During the 2011--2012 and 2012--2013 data taking runs, a significant
improvement to the systematic error in the determination of the
neutron polarization resulted from the installation of a physical
shutter in the region between the UCN decay trap and the upstream
guide region feeding the spectrometer \cite{brown18}.  An in-situ
measurement of the polarization was carried out on a run-by-run basis
following each $\beta$-decay run.  The important components to these
in-situ measurements of the polarization (see Fig.\ \ref{fig1}) were
the UCN decay volume, the shutter, the guide region feeding the
spectrometer, the spin flipper and 7 T polarizer, and then a UCN
detector which could be directly connected to the guide region via
a switcher.

\begin{figure}[h]
\centering
\includegraphics[scale=0.40]{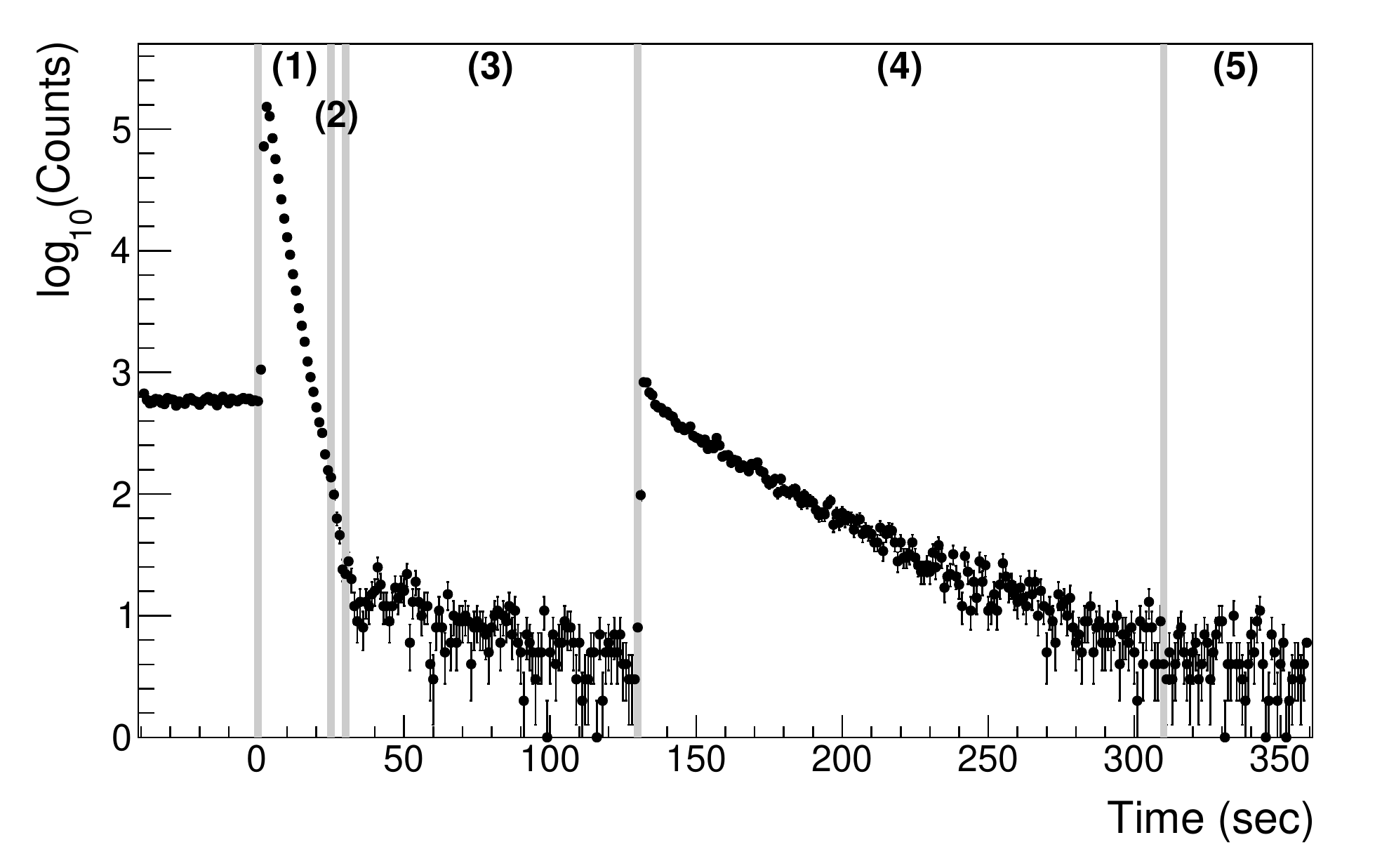}
\caption{Counts in the UCN detector during the depolarization
measurement.  See text for details.}
\label{fig2}
\end{figure}

A five step procedure for the measurement of the polarization was as
follows.  In Step (1), following each beta decay run, during which an
equilibrium spin state population had developed in the spectrometer
(i.e., with the spin flipper operated in some nominal state), the
shutter was closed, and the UCN detector was connected to the guide
via the switcher.  At this point, UCN were trapped by the shutter
within the decay volume, while UCN of the nominal spin state upstream
of the shutter were drained into the UCN detector.  The shutter
significantly improved the signal-to-noise ratio for the next steps.
In Step (2), the spin flipper state was then toggled from its nominal
state during $\beta$-decay running, so that depolarized UCN in the
guide region feeding the spectrometer, which were previously trapped
by the 7 T polarizing field, had their spins flipped so that they
could then transit the 7 T field and then drain into the detector.  In
Step (3), the shutter was then opened, and with the spin flipper still
in its toggled state, depolarized UCN in the decay trap could exit the
decay trap region, have their spins flipped permitting them to transit
the 7 T field, and then drain into the UCN detector.  In Step (4), the
spin flipper was reset back to its nominal state, which then permitted
a measurement of the polarized UCN within the decay trap.  Finally, in
Step (5), background data in the detector were taken.
Fig.\ \ref{fig2} shows an example of counts in the detector during the
five step procedure described above.

A series of Monte Carlo simulations were then performed to correct for
two effects: a so-called ``depolarization evolution'' effect, namely
that the depolarized population within the decay trap continued to be
fed while the shutter was closed, and also a correction for the finite
spin flipper efficiency.  To study this, simulations were carried out
at the NERSC (National Energy Research Scientific Computing Center)
facility, in which $\chi^2$ searches were performed by varying the
guide specularity, Fermi potential, etc.  It is important to point out
that the systematic error we quote on the polarization (see below) was
dominated by the statistical uncertainties in the fitting procedures
for the depolarization evolution and finite spin flipper efficiency
corrections (i.e., by the counting statistics in the detector).
Complete details on the depolarization measurement may be found in
Ref.\ \cite{dees-tobepublished}.

\section{Spectrometer Calibration}
\label{spectrometer_calibration}

\begin{figure}[h]
\centering
\includegraphics[scale=0.45]{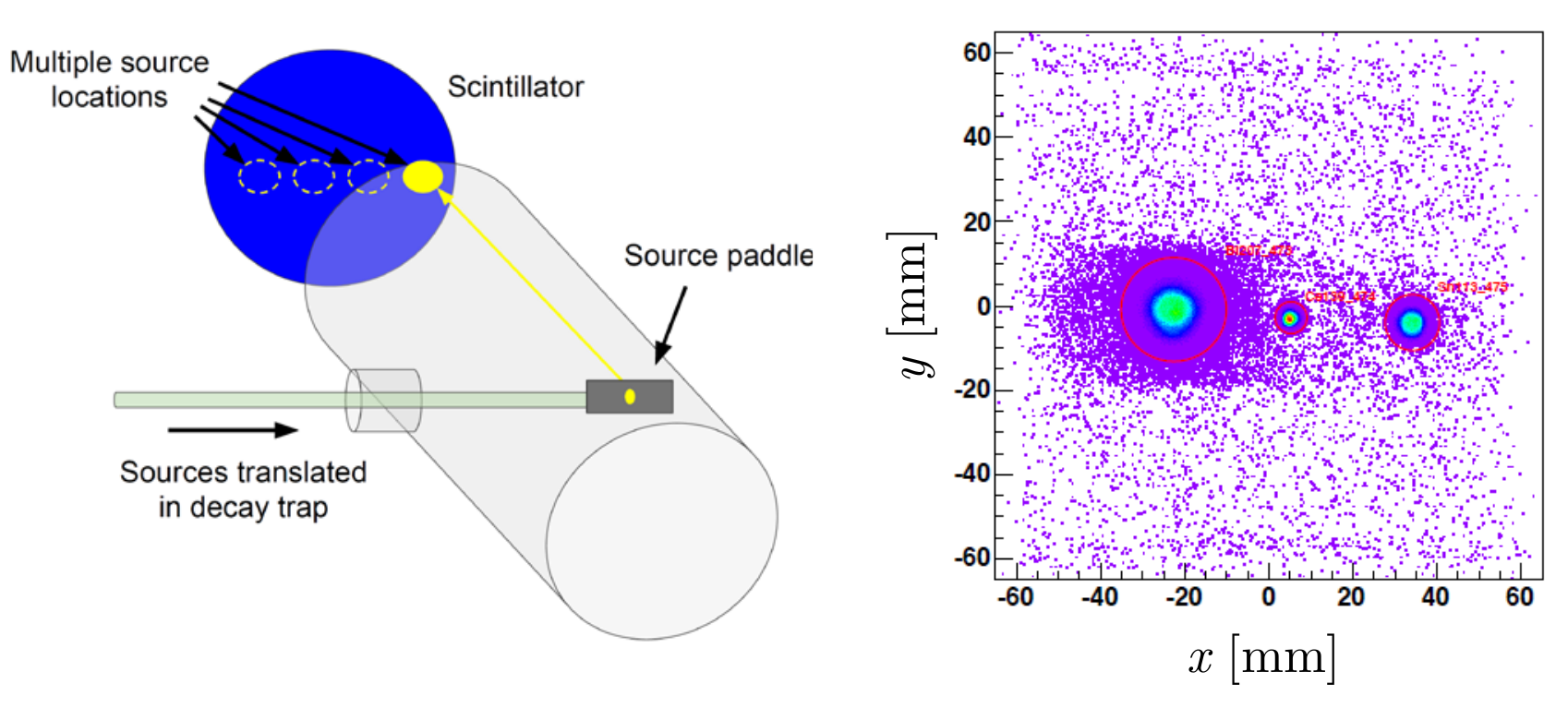}
\caption{Left: illustration of the ``load lock'' source insertion
system, which permitted sealed sources to be translated across the
detector face with the spectrometer under vacuum.  Right: example of a
reconstruction by the MWPC of the transverse $(x,y)$ positions of
three different calibration sources (left to right: $^{207}$Bi,
$^{139}$Ce, and $^{113}$Sn).}
\label{fig3}
\end{figure}

The electron detectors were calibrated using a ``load lock'' system
which permitted sealed sources to be translated into the decay trap,
and then scanned across the detector face, without breaking vacuum in
the spectrometer.  The location of this load lock insertion system can
be seen in Fig.\ \ref{fig1} (denoted ``Calibration Source
Insertion'').  Fig.\ \ref{fig3} then illustrates the translation of
the sealed sources across the detector, together with an example
reconstruction by the MWPC of the transverse $(x,y)$ positions of
three different calibration sources (left to right: $^{207}$Bi,
$^{139}$Ce, and $^{113}$Sn) within the spectrometer volume.

The thicknesses of the foils encapsulating these three sources were
measured in an offline setup using $\alpha$ particles from a
collimated $^{241}$Am source and a silicon detector.  A comparison of
the measured energy losses in these foils with Geant4 simulations
indicated the source foil thicknesses were 9.4 $\mu$m, in contrast
with the manufacturer's nominal specification of 7.6 $\mu$m
\cite{mendenhall-thesis}.  These 9.4 $\mu$m thicknesses were then
included in the simulations of the energy spectra from the calibration
sources.

Calibrations of the so-called ``visible energy'', $E_{\mathrm{vis}}$,
for each PMT $i$ were carried out according to a model
\cite{mendenhall-thesis, brown-thesis} in which
\begin{equation}
E_{\mathrm{vis},i} = \eta_i^{-1}(x,y) \cdot
f_i\left(\left(\mathrm{ADC}_i - p_i(t)\right)\cdot g_i(t) \right),
\end{equation}
where $\eta_i^{-1}(x,y)$ denotes an unfolding of the $(x,y)$-position
dependent response of the system (i.e., due to the position-dependent
light response of the system), and the function
$f_i\left(\left(\mathrm{ADC}_i - p_i(t)\right)\cdot g_i(t) \right)$
represents the linearity of the system between light output in the
scintillator and the ADC channel ultimately read-out by the data
acquisition system for each PMT signal, with $p_i(t)$ and $g_i(t)$
representing the time-dependent pedestal and gain correction factor,
respectively.  Values for $\eta_i(x,y)$ in multiple $(x,y)$ bins were
obtained from special calibration runs carried out with activated
xenon gas which uniformly filled the decay trap volume; the
calibration especially utilized the endpoint of $^{135}$Xe decays,
whereby the fitted endpoints in each $(x,y)$ bin were compared with a
fixed value, thus providing for a relative $\eta_i(x,y)$ ``position
map''.  The linearity was obtained from a comparison of the system
response to the sources at multiple $(x,y)$ positions, via a
comparison of the simulated visible energy multiplied by the
$\eta_i(x,y)$ position map obtained from the xenon calibration data,
with the observed ADC channel number.

\begin{figure}
\centering
\includegraphics[scale=0.50]{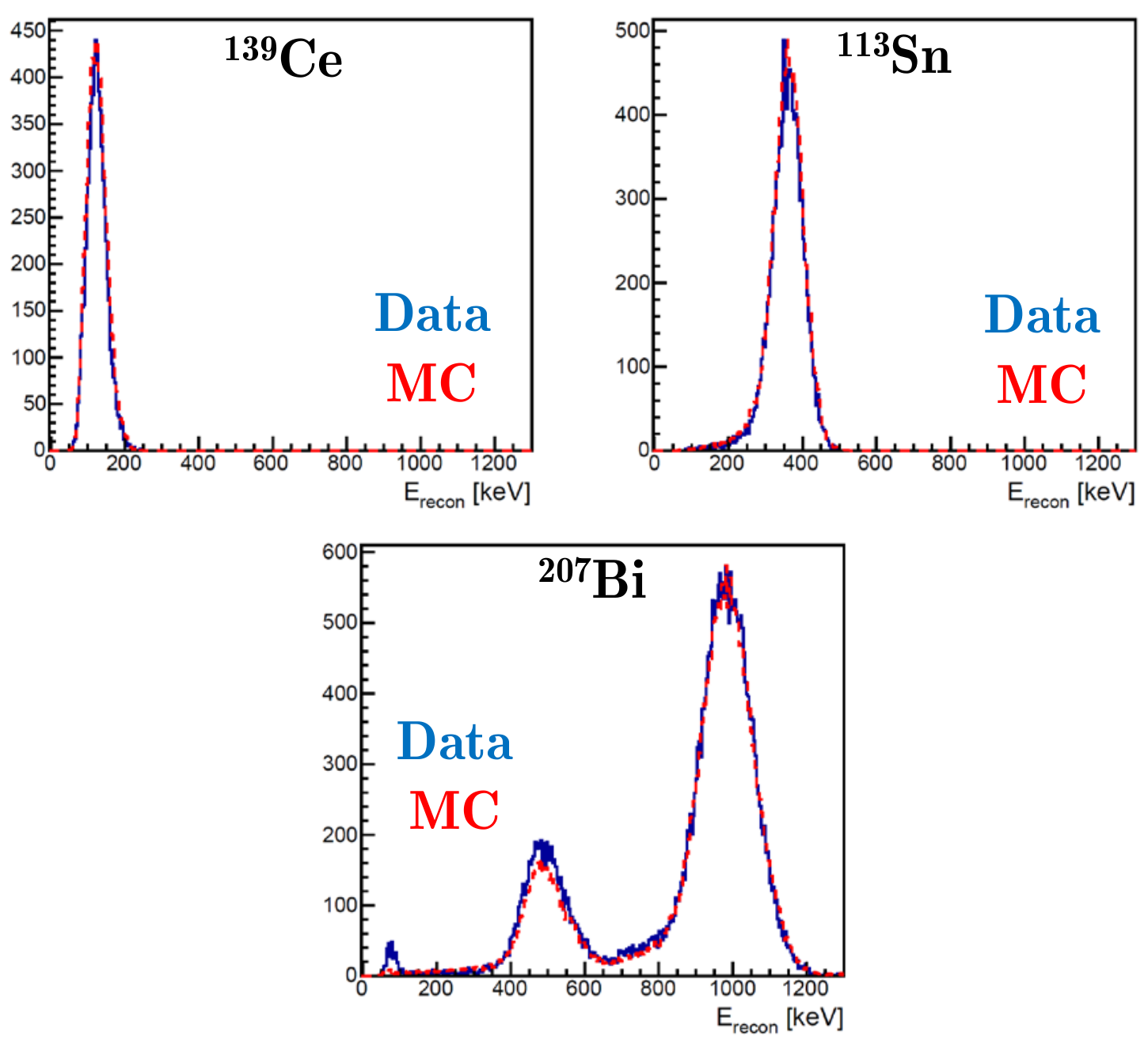}
\caption{Comparison of simulated (red) and data (blue) spectra from
calibration runs with $^{139}$Ce, $^{113}$Sn, and $^{207}$Bi
calibration sources.}
\label{fig4}
\end{figure}

A verification of the efficacy of the calibration can be seen in
Fig.\ \ref{fig4}, which compares simulated and reconstructed (i.e.,
calibrated) spectra for runs with $^{139}$Ce, $^{113}$Sn, and
$^{207}$Bi calibration sources.  The agreement between simulation
and data is seen to be quite good.

\section{Event Types, Backscattering, and $\langle\cos\theta\rangle$
Corrections}

\begin{figure}
\centering
\includegraphics[scale=0.45]{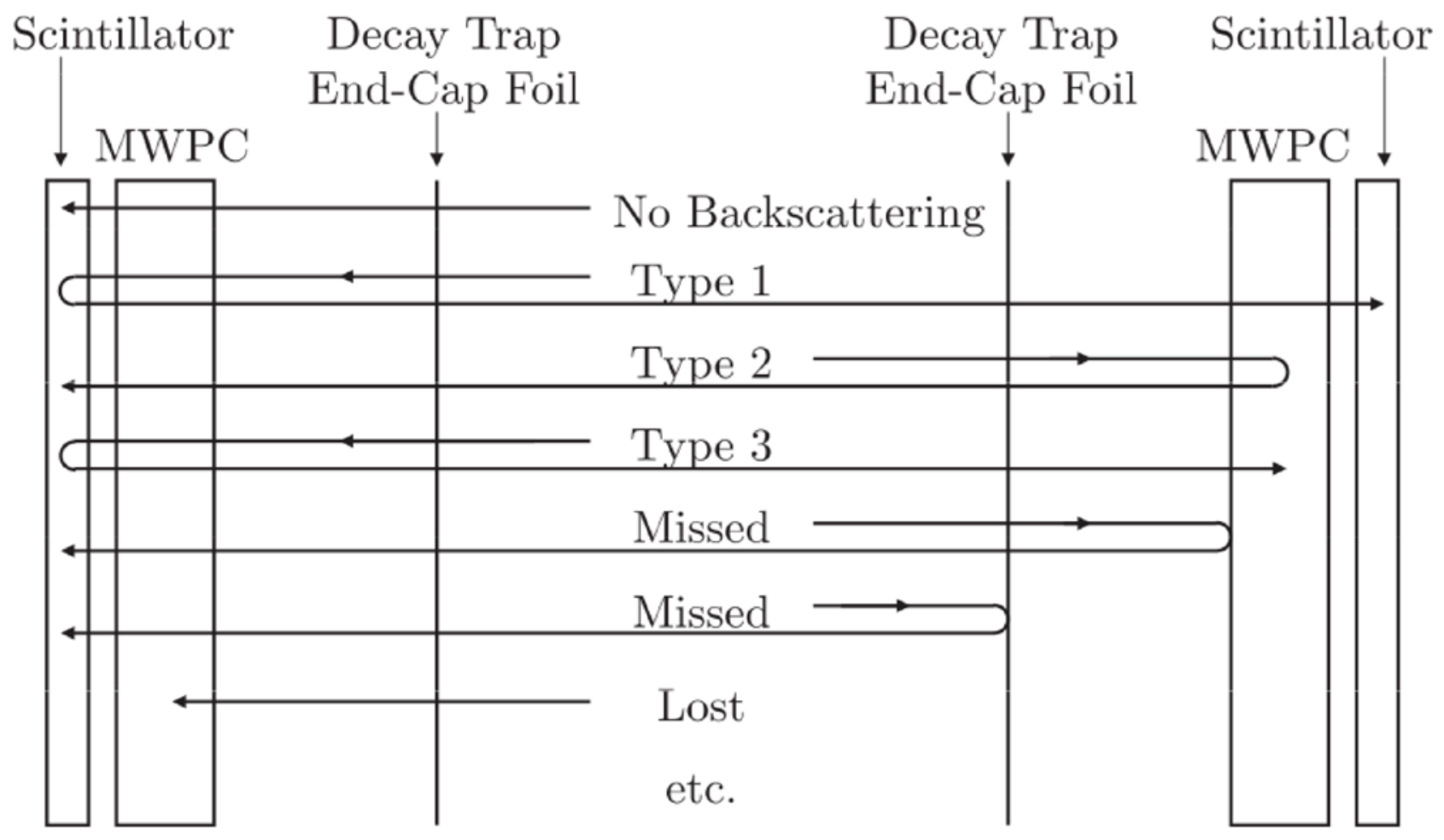}
\caption{Classification of the different types of events in the
experiment.}
\label{fig5}
\end{figure}

The measurement of the asymmetry requires corrections for a number of
different types of backscattering events.  Fig.\ \ref{fig5}
illustrates a classification of the different types of events in the
experiment, indicating whether a signal was recorded in each
scintillator and/or MWPC.  We define ``Type 0'' events (i.e., those
in which there was no \textit{reconstructable} backscattering) to
be the sum of ``No Backscattering'' events plus ``Missed Backscattering''
events.

Corrections for each event type were calculated using independent
Geant4 and PENELOPE simulations; the agreement between the simulations
was shown to be quite good.  We define two types of
corrections, each defined such that the corrected asymmetry is
$|A_\mathrm{corr}| \equiv |A_\mathrm{uncorr}|(1+\Delta)$ (i.e.,
a positive correction $\Delta > 0$ indicates that the magnitude
of the asymmetry would be increased).  The first is a correction
we term $\Delta_2$, where $\Delta_2$ corrects for missed and incorrectly
identified backscattering.  These types of events would otherwise
dilute the asymmetry; therefore, these corrections are
expected to be positive.  The second is a correction we term
$\Delta_3$, which we term a $\langle\cos\theta\rangle$ correction.
This is so named to account for the deviation of $\langle\cos\theta\rangle$
from a value of 1/2 over each hemisphere, due to the angular-dependent
acceptance of the spectrometer and detectors.  Because low pitch angle
(i.e., large $\cos\theta$) and high-energy events are more likely
to be detected, this results in a positive bias to the magnitude
of the measured asymmetry; therefore, the values of $\Delta_3$
are expected to be negative in order to remove this bias.

\begin{figure}
\centering
\includegraphics[scale=0.43]{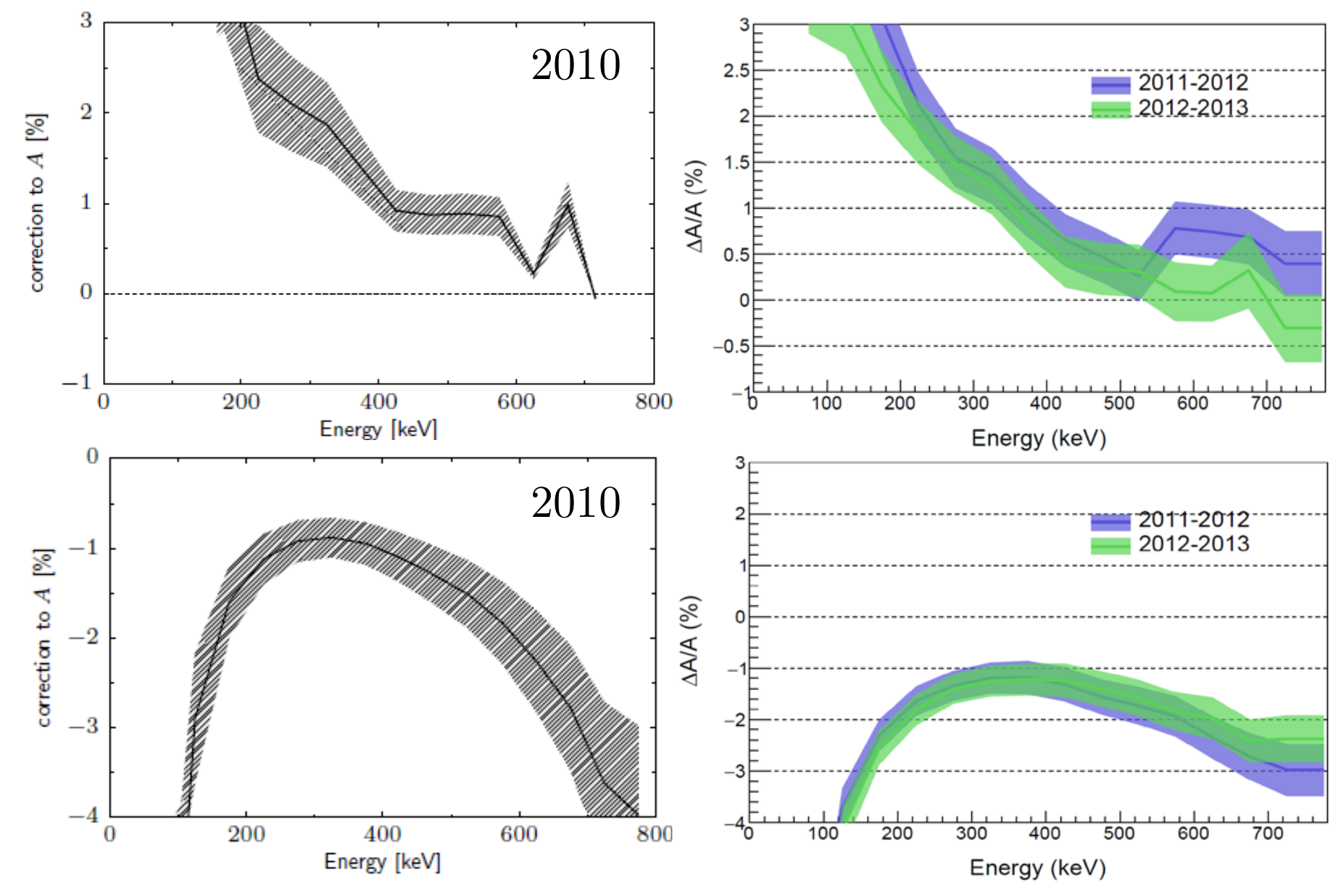}
\caption{Calculated values of the $\Delta_2$ (backscattering, top panels)
and $\Delta_3$ ($\langle\cos\theta\rangle$, bottom panels) corrections
as a function of the electron energy for the 2010 (left panels) and
2011--2012 and 2012--2013 data sets (right panels).}
\label{fig6}
\end{figure}

Calculated values for the $\Delta_2$ and $\Delta_3$ corrections are
shown as a function of the electron energy in Fig.\ \ref{fig6} for the
2010, 2011--2012, and 2012--2013 data sets.  As expected, the
magnitude of the corrections decreased as the decay trap and MWPC foil
thicknesses progressively decreased with each data set.

\begin{table*}
\caption{Summary of the systematic and statistical errors for the
2010, 2011--2012, and 2012--2013 data taking runs.  Note that a $+$
($-$) sign correction indicates the corrections increases
(decreases) the magnitude of the asymmetry.}  \centering
\begin{tabular}{|l|cc|ccc|} \hline\hline
& \% Corr.& \% Unc.& \multicolumn{2}{c}{\% Corr.}&
  \% Unc. \\
Effect& 2010& 2010& 2011--2012& 2012--2013& 2011--2013 \\ \hline
Polarization&   $+0.67$& $\pm 0.56$& $+0.45$& $+0.34$& $\pm 0.17$ \\
Backscattering& $+1.36$& $\pm 0.34$& $+1.08$& $+0.88$& $\pm 0.30$ \\
$\langle\cos\theta\rangle$& $-1.21$& $\pm 0.30$& $-1.53$& $-1.51$&
  $\pm 0.33$ \\
Energy Reconstruction& ---& $\pm 0.31$& ---& ---& $\pm 0.20$ \\
Gain Fluctuation& ---& $\pm 0.18$& ---& ---& $\pm 0.16$ \\
Field Nonuniformity& $+0.06$& $\pm 0.10$& ---& ---& $\pm 0.12$ \\
Muon Veto Efficiency& ---& $\pm 0.03$& ---& ---& $\pm 0.03$ \\
UCN-Induced Background& $+0.01$& $\pm 0.02$& $+0.01$& $+0.01$& $\pm 0.02$ \\
MWPC Efficiency& $+0.12$& $\pm 0.08$& $+0.13$& $+0.11$& $\pm 0.01$ \\ \hline
Total Systematics& & $\pm 0.82$& & & $\pm 0.52$ \\ \hline
Statistics& & $\pm 0.46$& & & $\pm 0.36$ \\ \hline
Recoil Order Effects \cite{bilenkii60,holstein74,wilkinson82,gardner01}&
  $-1.71$& $\pm 0.03$& $-1.68$& $-1.67$& $\pm 0.03$ \\
Radiative Effects \cite{shann71, gluck92}&
  $-0.10$& $\pm 0.05$& $-0.12$& $-0.12$& $\pm 0.05$ \\ \hline\hline
\end{tabular}
\label{tab:error_budget}
\end{table*}

\section{Error Budgets}

A summary of the error budgets for the 2010 \cite{mendenhall13},
2011--2012 \cite{brown18}, and 2012--2013 \cite{brown18} data sets is
shown in Table \ref{tab:error_budget}.  As already noted above, the
significant decrease in the systematic error associated with the
polarization resulted from the installation of the shutter in between
the 2010 and 2011--2012 data taking runs.  Ultimately, as can be seen
in the table, the reach of the experiment was limited by the
systematic uncertainties in the corrections for backscattering and the
$\langle\cos\theta\rangle$ acceptance, both of which were on the scale
of the statistical error bar.  A future UCNA+ experiment
will need to be designed such that these effects are significantly reduced
in order for a $<0.2$\% precision to be obtained on the asymmetry.

\section{Summary of UCNA Results for $A$}

A summary of all of the UCNA results for $A$ is given in
Table \ref{tab:results}.  
The final result from the combination of the data sets obtained during 2010
\cite{mendenhall13} and 2011--2013 \cite{brown18} is
$A_0 = -0.12015(34)_\mathrm{stat}(63)_\mathrm{syst}$.

\begin{table}
\caption{Summary of all results from the UCNA experiment.}
\begin{tabular}{cccc} \hline\hline
Data& & & Year \\
Set&
  $\delta A/A_\mathrm{stat}$ [\%]& $\delta A/A_\mathrm{syst}$ [\%]&
  Published \\ \hline
2007& 4.0& 1.8& 2009 \cite{pattie09} \\
2008--2009& 0.74& 1.1& 2010 \cite{liu10, plaster12} \\
2010& 0.46& 0.82& 2013 \cite{mendenhall13} \\
2011--2013& 0.37& 0.56& 2018 \cite{brown18} \\ \hline
\end{tabular}
\label{tab:results}
\end{table}

\section{Impact of the UCNA Experiment}

\begin{figure}
\centering
\includegraphics[scale=0.42]{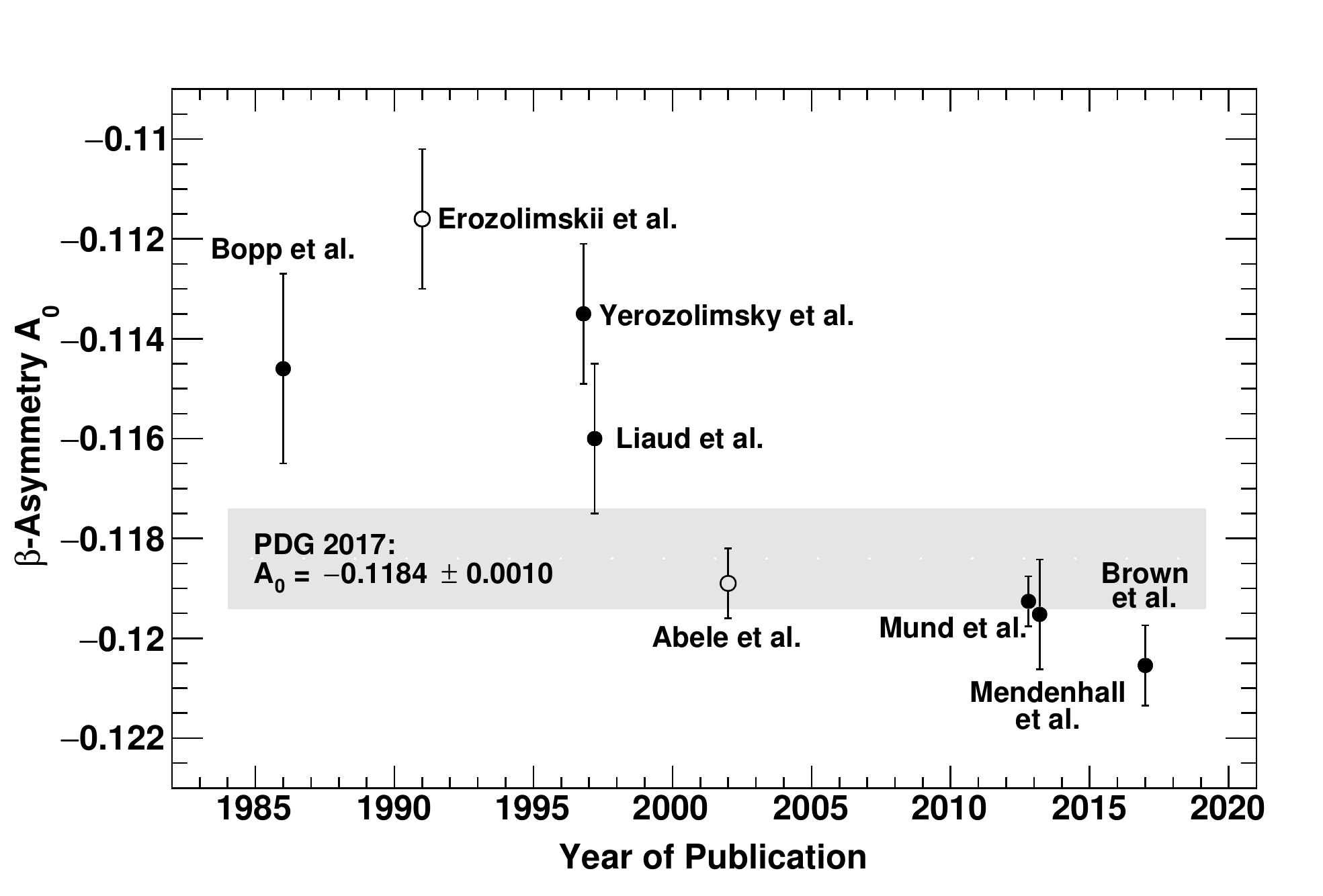}
\caption{Results for $A$ \cite{bopp86, yerozolimsky97, liaud97,
abele02, mund13, mendenhall13, brown18} plotted vs.\ year of
publication.}
\label{fig7}
\end{figure}

With the UCNA experiment now concluded, the long-term impact of our
final result can be seen in Fig.\ \ref{fig7}.  There, one can see the
striking landscape of the time evolution of values for $A$
\cite{bopp86, yerozolimsky97, liaud97, abele02, mund13, mendenhall13,
brown18}, shown plotted vs.\ publication year.  It should be noted
that the $\sqrt{\chi^2/\nu}$ scale factor the Particle Data Group
\cite{PDG} applies to the error is rather large, $\sim 2.4$, due to
the rather striking dichotomy between many of the older and more
recent values.
A common theme that emerges between many of the older and more recent
results concerns the size of the systematic corrections.  Generally
speaking, in many of the older results, the systematic corrections
were of the order of $>2$\%, whereas in the more recent results, the
corrections were all of the order of $<2$\%.

In preparing our most recent publication \cite{brown18}, we discovered
that the PDG only includes in the calculation of the scale factor
those measurements that satisfy $\delta x_i < 3 \sqrt{N} \delta
\bar{x}$, where $x_i$ refers to one measurement of quantity $x$ out of
$N$ measurements and $\delta \bar{x}$ is the non-scaled error on the
weighted average $\bar{x}$ \cite{PDG}.  Inclusion of a 0.1\% result
for $A_0$ would remove many of the older results for $A$ from those
that enter the calculation of the scale factor.  With the expected
forthcoming results from the PERKEO III experiment, this could be a
real turning point in progress for the field, whereby the PDG may
potentially no longer need to apply a $\sqrt{\chi^2/\nu}$ scale factor
to the average value of $A$.

\section{Acknowledgments}
This work was supported in part by the U.S.\ Department
of Energy, Office of Nuclear Physics
(DE-FG02-08ER41557, DE-SC0014622, DE-FG02-97ER41042) and the National
Science Foundation (NSF-0700491, NSF-1002814, NSF-1005233, NSF-1102511,
NSF-1205977, NSF-1306997,
NSF-1307426, NSF-1506459,
and NSF-1615153). We gratefully acknowledge the support
of the LDRD program (20110043DR), and the LANSCE
and AOT divisions of the Los Alamos National Laboratory.

We thank the organizers of the PPNS-2018 workshop for selecting this abstract
for an oral presentation, and for their excellent hospitality during this
outstanding decennial workshop.

\end{document}